\documentclass[10pt,letterpaper]{article}

\usepackage[top=0.85in,left=2.in,footskip=0.75in]{geometry}

\usepackage{amsmath,amssymb}

\usepackage{graphicx}
\usepackage{epstopdf}

\begin{document}

{\Large
		\noindent	\textbf{Multi-stage models for the failure of complex systems, cascading disasters, 
				and the onset of disease} 
}
\\

\noindent
Anthony J. Webster
\\

\noindent
Nuffield Department of Population Health, Richard Doll
  Building, University of Oxford, Old Road Campus, Oxford, OX3 7LF,
  UK.
\\

\noindent
E-mail: anthony.webster@ndph.ox.ac.uk

\section*{Abstract}
Complex systems can fail through different routes, often progressing through a series 
of (rate-limiting) steps and modified by environmental exposures.  
The onset of disease, cancer in particular, is no different. 
Multi-stage models provide a 
simple but very general mathematical framework for studying 
the failure of complex systems, or equivalently, the onset of disease.  
They include the Armitage-Doll multi-stage cancer model as a particular case, and
have potential to provide new insights into how failures and disease, arise and progress. 
A method described by E.T. Jaynes is developed to provide an analytical 
solution for a large class of these
models, and highlights connections between the convolution 
of Laplace transforms, sums of random variables, and Schwinger/Feynman parameterisations. 
Examples include: exact solutions to the Armitage-Doll model, the sum of Gamma-distributed 
variables with integer-valued shape parameters, a clonal-growth cancer model, and a 
model for cascading disasters. 
Applications and limitations of the approach  
are discussed in the context of recent cancer research. 
The model is sufficiently general to be used in many contexts, such as engineering, 
project management, disease progression, and disaster risk for example, allowing the 
estimation of failure rates in complex systems and projects. 
The intended result is a mathematical toolkit for applying 
multi-stage models to the study of failure rates in 
complex systems and to the onset of disease, cancer in particular. 



\section{Introduction}

Complex systems such as a car can fail through many different routes, often 
requiring a sequence or combination of events for a component to fail.
The same can be true for human disease, cancer in particular \cite{Nordling1953,DollArmitage,Peto2016}. 
For example, cancer can arise through a sequence of steps such as genetic 
mutations, each of which must occur prior to cancer 
\cite{Moolgavkar2016, LuebeckMoolgavkar2002, Michor2006, MezaLuebeck2008, Little2010}.  
The considerable genetic variation between otherwise similar cancers 
\cite{Vogelstein2013,Martincorena2015}, suggests that similar cancers might arise through a 
variety of different paths.    
Multi-stage models describe how systems can fail through 
one or more possible routes. 
They are sometimes described as ``multi-step'' or ``multi-hit'' 
models \cite{Ashley1969, Knudson1971}, because each  
route typically requires failure of one or more 
sequential or non-sequential steps. 
Here we show that the model is easy to conceptualise and derive, and that 
many specific examples have analytical solutions or approximations, making it
ideally suited to the construction of biologically- or
physically-motivated models for the incidence of events such as
diseases, disasters, or mechanical failures. 
A method described by E.T. Jaynes \cite{Jaynes} generalises to give an exact analytical 
formula for the sums of random variables needed to evaluate the sequential model. 
This is evaluated for specific cases. 
Moolgavkar's exact solution \cite{Moolgavkar} to the Armitage-Doll multistage cancer
model is one example that is derived surprisingly easily, and is easily modified.   
The approach described here can incorporate simple models for a clonal expansion prior to cancer detection \cite{LuebeckMoolgavkar2002, Michor2006, MezaLuebeck2008}, but as discussed in Sections \ref{cascading} and \ref{evolution}, it may not be able to describe evolutionary competition or  
cancer-evolution in a changing micro-environment without additional modification. 
More generally, it is hoped that the mathematical framework can be used in a broad 
range of applications, including the modelling of other 
diseases \cite{Al-Chalabi2014, Chio2018, Corcia2018, Licher2019}. 
One example we briefly describe in Section \ref{cascading} is modelling 
of ``cascading disasters'' \cite{AghaKouchak}, where each disaster 
can substantially modify the risk of subsequent (possibly different) disasters. 
Conventional notation is used \cite{Collett}, with: probability densities $f(t)$,
cumulative probability distributions $F(t)=\int_0^t f(t)$, a 
survival function $S(t)=1-F(t)$, hazard function $h(t)=f(t)/S(t)$,
and cumulative hazard function $H(t)= \int_0^t h(y)dy$. 
Noting that $f(t)=-dS/dt$, it is easily seen that $H(t)=\int_0^t
f(y)/S(y) dy=-\log S(t)$, $h(t)=-d\log S(t)/dt$, and $S(t)=\exp(
-\int_0^t h(y)dy )$. 

\section{Failure by multiple possible routes}

\begin{figure}[ht]
  \centering{\includegraphics[width=0.5\linewidth]{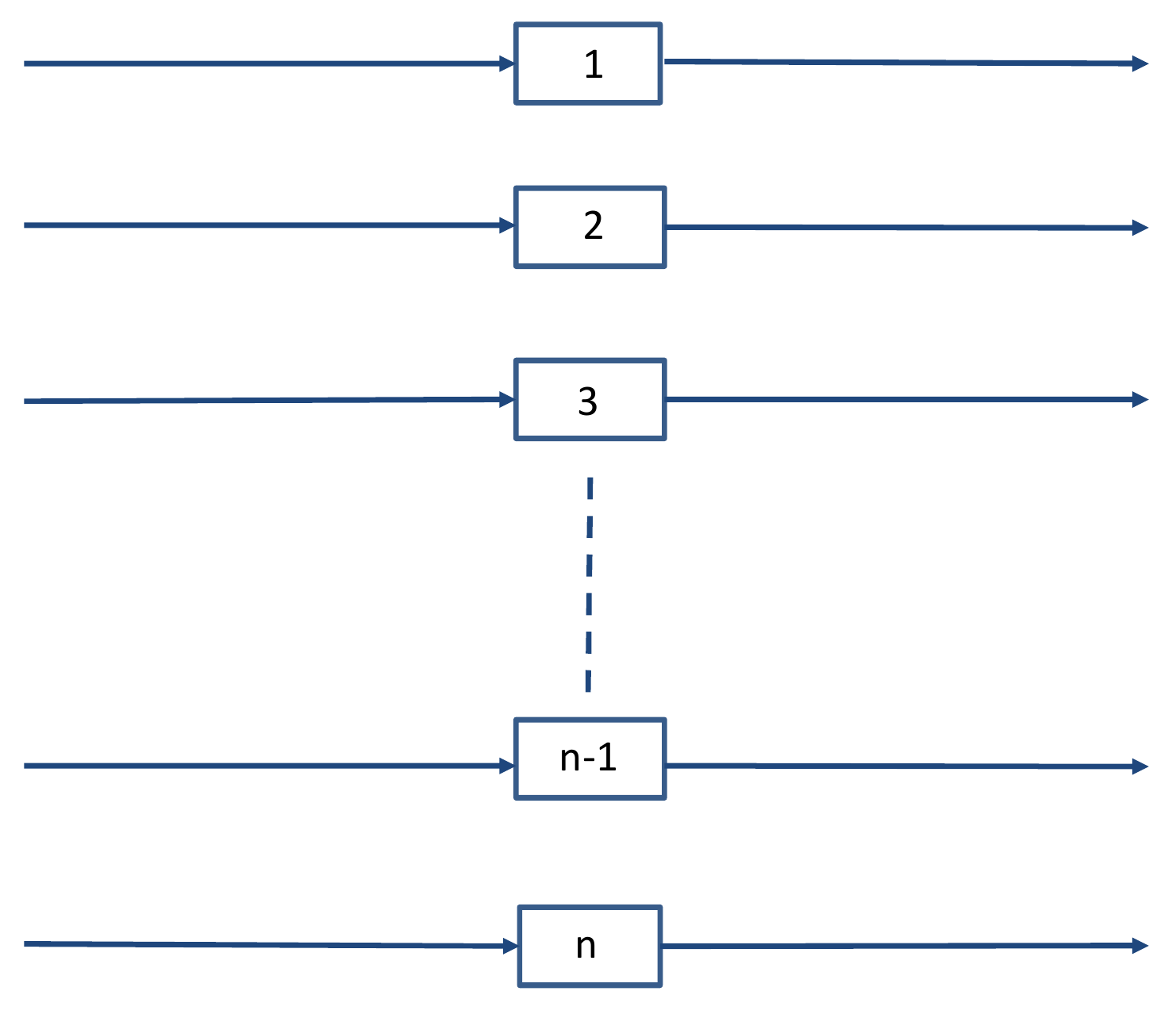}}
  \vspace{0.4cm}
  \caption{\sl In a complex system, failure can occur through many different routes (Eq. \ref{S}).}
  \label{fig1}
\end{figure}

Imagine that we can enumerate all possible routes $1$ to $n$ by which
a failure can occur (Fig \ref{fig1}).  
The probability of surviving the $i$th of these routes after time $t$
is $S_i(t)$, and consequently the probability of surviving all of
these possible routes to failure $S(t)$ is, 
\begin{equation}\label{S}
S(t) = \Pi_{i=1}^n S_i(t)
\end{equation}
or in terms of cumulative hazard functions with $S_i(t)=e^{-H_i(t)}$, 
\begin{equation}\label{SH}
S(t) = \exp \left\{ - \sum_{i=1}^n H_i(t) \right\} 
\end{equation}
The system's hazard rate for failure by any of the routes is, 
\begin{equation}
\begin{array}{ll}
h(t) &= -\frac{d}{dt} \log \left( S(t)   \right)
\\   &= - \sum_{i=1}^n \frac{d}{dt} \log \left( S_i(t) \right) 
\\   &= \sum_{i=1}^n h_i(t)
\end{array}
\end{equation}
and $H(t)=\sum_{i=1}^n H_i(t)$.
In words, if failure can occur by any of $n$ possible routes, the
overall hazard of failure equals the sum of the hazard of failure by
all the individual routes.  

A few notes on Eq. \ref{SH} and its application to cancer modelling. 
Firstly, if the $s$th route to failure is much more likely than the others,
with $H_{s} \gg H_j$ for  $s\neq j$, then 
$S(t) = \exp\left\{ 
-H_s(t) + \left( 1 + O \left( \sum_{i\neq s} H_i/H_s \right) \right) 
\right\} 
\simeq \exp \left\{ - H_s(t) \right\}$,   
which could represent the most likely sequence of mutations in a cancer model for example. 
Due to different manufacturing processes, genetic backgrounds, chance processes or exposures 
(e.g. prior to adulthood), this most 
probable route to failure could differ between individuals. 
Secondly, the stem cell cancer model assumes that cancer can occur through any of $n_s$ 
equivalent stem cells in a tissue, for which 
Eq. \ref{SH} is modified to,
$S= \exp \left\{ - n_s \sum_{i=1}^n H_i(t) \right\}$.  
So a greater number of stem cells is expected to increase cancer risk, 
as is observed \cite{Tomasetti2015, Nunney2018}.  
Thirdly, most cancers are sufficiently rare that $S \sim 1$. 
As a consequence, many cancer models (implicity or explicitly) 
assume $S\simeq 1-n_s\sum_{i=1}^n H_i(t)$ and 
$f=-dS/dt\simeq n_s \sum_{i=1}^n h_i(t)$,  
a limit 
emphasised in 
the Appendix of Moolgavkar \cite{Moolgavkar}. 

\section{Failure requiring {\it m} independent events}\label{mIndependent}

\begin{figure}[ht]
  \centering{\includegraphics[width=0.5\linewidth]{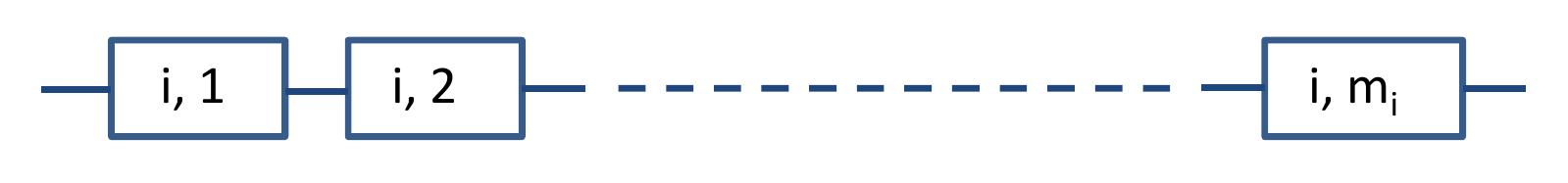}}
  \vspace{0.4cm}
  \caption{\sl Failure by the $i$th path at time $t$ requires $m_i$ independent failures to 
  occur in any order, with the last failure at time $t$ (Eq. \ref{S_i_indp}).}
  \label{fig2}
\end{figure}

Often failure by a particular path will require more than one failure
to occur independently. 
Consider firstly when there are $m_i$ steps to failure, and the order
of failure is unimportant (Fig \ref{fig2}).  
The probability of surviving failure by the $i$th route, $S_i(t)$ is, 
\begin{equation}\label{S_i}
\begin{array}{ll}
S_i(t) &= P \left( \mbox{survive any one or more, necessary step for failure} \right)
\\     &= 1 - P \left( \mbox{fail all the steps} \right)
\\     &= 1 - \Pi_{j=1}^{m_i} F_{ij}(t) 
\end{array}
\end{equation}
where $F_{ij}(t)$ is the cumulative probability distribution for failure of the
$j$th step on the $i$th route within time $t$. 
Writing $S_{ij}(t)=1-F_{ij}(t)$, this can alternately be written as,
\begin{equation}\label{S_i_indp}
S_i(t) = 1 - \Pi_{j=1}^{m_i} \left( 1 - S_{ij}(t) \right) 
\end{equation}  

\section{Relation to recent multi-stage cancer models}

It may be helpful to explain how Eqs. \ref{S} and \ref{S_i} 
are used in recently described multi-stage cancer models 
\cite{Wu, Zhang, Calabrese}. 
If we take a rate of mutations $\mu_j$ per cell division for each of the rate-limiting 
mutational steps $1$ to $j$, and $d_i$ divisions of cell $i$, then the probability of a 
stem cell surviving without the $j$th rate limiting mutation is $S_{ij}=(1-\mu_j)^{d_i}$. 
Similarly, the probability of a given stem cell having mutation $j$ is 
$F_{ij}=1-(1-\mu_j)^{d_i}$. 
This is the solution of Zhang et al. \cite{Zhang} to the recursive formula of 
Wu et al. \cite{Wu} (see Appendix of Zhang et al. \cite{Zhang} for details). 
Using Eq. \ref{S_i}, the survival of the $i$th stem cell is described by,
\begin{equation}
S_i = 1 - \Pi_{j=1}^{m_i} \left( 1 - \left( 1-\mu_j \right)^{d_i} \right) 
\end{equation}
Now assuming all $n$ stem cells are equivalent and have equal rates $\mu_i=\mu_j$ 
for all $i$, $j$, and consider only one path to cancer with $m$ mutational steps, then,  
\begin{equation}
S_i = 1 - \left( 1 - \left( 1- \mu \right)^d \right)^m
\end{equation}
and, 
\begin{equation}\label{S-Model}
\begin{array}{ll}
S 
&= \Pi_{i=1}^n S_i
\\
&=\left( 1 - \left( 1 - (1-\mu)^d \right)^m \right)^n 
\end{array}
\end{equation}
The probability of cancer within $m$ divisions, often referred to as ``theoretical lifetime 
intrinsic cancer risk", is, 
\begin{equation}\label{F}
F=1 - \left( 1 - \left( 1 - (1-\mu)^d \right)^m \right)^n 
\end{equation}
This is the equation derived by Calabrese and Shibata \cite{Calabrese}, and that 
Zhang found as the solution to the model of Wu et al \cite{Zhang, Wu}. 
Therefore, in addition to the models of Wu and Calabrese being equivalent cancer models 
needing $m$ mutational steps, the models also assume that the order of the steps is not 
important. 
This differs from the original Armitage-Doll model that considered a sequential set of 
rate-limiting steps, and was exactly solved by Moolgavkar \cite{Moolgavkar}. 
Eqs. \ref{S-Model} and \ref{F} are equivalent to assuming: 
(i) equivalent stem cells, 
(ii) a single path to cancer, 
(iii) equivalent divisions per stem cell, and, 
(iv) equivalent mutation rates for all steps. 
Despite the differences in modelling assumptions for 
Eq. \ref{F} and the Armitage-Doll model, their predictions 
can be quantitatively similar. 
To see this, use the Armitage-Doll approximation of $\mu d \ll 1$, to expand,  
\begin{equation}
\left( 1 - \mu \right)^d = \exp\left( d\log(1-\mu) \right) \simeq \exp \left( \mu d\right) 
\end{equation}
If cell divisions are approximately uniform in time, then we can replace $\mu d$ with $\mu t$, 
with $\mu$ now a rate per unit time. 
Then expanding $\exp(-\mu t)\simeq 1- \mu t$, gives,
\begin{equation}
F = 1 - \left( 1 - \left( 1 - \left( 1- \mu \right)^d \right)^m \right)^{n_s}
\simeq 1 - 
\left( 1 - \left( \mu t \right)^m \right)^{n_s}  
\simeq n_s \left( \mu t \right)^m 
\end{equation}
The incidence rate $h = f/S$ is then $h \simeq n_s \mu^m t^{m-1}$, the same as 
the original (approximate) Armitage-Doll  solution \cite{DollArmitage}. 
This approximate solution is expected to become inaccurate at sufficiently long times. 
An equivalent expression to Eq. \ref{S-Model} was known to Armitage, Doll, and Pike 
since at least 1965 \cite{Pike}, as was its limiting behaviour for large $n$. 
The authors \cite{Pike} emphasised that many different forms for the $F_i(t_i)$ could produce 
approximately the same observed $F(t)$, especially for large $n$, with the behaviour 
of $F(t)$ being dominated by the small $t$ behaviour of $F_i(t)$. 
As a result, for sufficiently small times power-law behaviour for $F(t)$ is likely, 
and if longer times were observable then an extreme value distribution would be 
expected \cite{Pike, Brody2012, Moolgavkar2016}. 
However the power-law approximation can fail for important cases with extra rate-limiting 
steps such as a clonal expansion \cite{LuebeckMoolgavkar2002, Michor2006, MezaLuebeck2008}. 
It seems likely that a model that includes clonal expansion and cancer detection is  
needed for cancer modelling, but the power law approximation could be used for 
all but the penultimate step, for example. 
A general methodology that includes this approach is described next, and examples are given 
in the subsequent section \ref{examples}. 
The results and examples of sections \ref{order} and \ref{examples} are 
intended to have a broad range of applications.

\section{Failure requiring {\it m} sequential steps}\label{order}

\begin{figure}[ht]
  \centering{\includegraphics[width=0.5\linewidth]{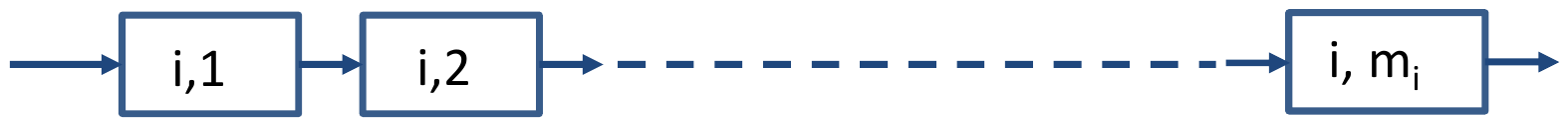}}
  \vspace{0.4cm}
    \caption{\sl Failure by the $i$th path at time $t$ requires an ordered sequence of failures, 
  with the last failure at time $t$ (Eqs. \ref{analyticSolution} and \ref{seqSi}).}
  \label{fig3}
\end{figure}

Some failures require a {\sl sequence} of independent events to occur, each following the one 
before (Fig \ref{fig3}). 
A well-known example is the Armitage-Doll multistage cancer model, that requires a 
sequence of $m$ mutations (failures), that each occur with a different constant rate. 
The probability density for failure time is the pdf for a sum of the 
$m$ independent times $t_j$ to failure at each step in the sequence, each of which may have 
a different probability density function $f_j(t_j)$. 
A general method for evaluating the probability density is outlined below, adapting a 
method described by Jaynes \cite{Jaynes} (page 569). 
Take $T_i \sim f_i(t_i)$ as random variables. 
Then use marginalisation to write $P \left( \sum_{j=1}^{m} T_j = t\right)$ in 
terms of $P \left( \sum_{j=1}^m T_j = t, T_1=t_1, \dots , T_m=t_m \right)$, where 
$(A,B,C)$ is read as ``$A$ and $B$ and $C$", and expand using the product 
rule $P(A,B)=P(A|B)P(B)$,
\begin{equation}
\begin{array}{lll}
P\left( \sum_{j=1}^m T_j = t \right) &= 
\int_0^{\infty} dt_1 \dots \int_0^{\infty} dt_m 
&P \left( \sum_{j=1}^m T_j = t, T_1=t_1, \dots , T_m=t_m \right)
\\
&= \int_0^{\infty} dt_1 \dots \int_0^{\infty} dt_m 
&P \left( \sum_{j=1}^m  T_j = t | T_1=t_1, \dots , T_m=t_m \right)
\\
&& \times P \left( T_1=t_1, \dots , T_m=t_m \right)
\end{array}
\end{equation}
Noting that $P\left( \sum_{j=1}^m  T_j = t | T_1=t_1, \dots , T_m=t_m \right)$ 
is zero for $t\neq \sum_{j=1}^m t_j$ and  
$1=\int_0^{\infty} dt P \left( \sum_{j=1}^m  T_j = t | T_1=t_1, \dots , T_m=t_m \right)$, 
indicates that it is identical to a Dirac delta function 
$\delta \left( t-\sum_{j=1}^m  t_j \right)$. 
For independent events $P(T_1=t_1, \dots , T_m=t_m)=\Pi_{j=1}^m f_j(t_j)$ where 
$f_j(t_j) \equiv P_j(T_j=t_j)$.
Writing $f(t) \equiv P\left( \sum_{j=1}^{m} T_{j} = t \right)$, then gives, 
\begin{equation}\label{EqToTransform}
f(t) 
= 
\int_0^{\infty} dt_{1} \dots 
\int_0^{\infty} dt_{m} 
\Pi_{j=1}^{m} f_j(t_j) 
\delta \left( t - \sum_{j=1}^{m} t_{j} \right) 
\end{equation}
To evaluate the integrals, take the Laplace transform with respect to $t$,
to give, 
\begin{equation}\label{LEqToTransform}
\mathcal{L} [f]  = 
\int_0^{\infty} e^{-st} f(t) dt = 
\int_0^{\infty} dt_{1} \dots 
\int_0^{\infty} dt_{m} 
\Pi_{j=1}^{m} f_{j} \left( t_{j} \right) 
e^{-s\left( t_{1}+ \dots +  t_{m} \right)}
\end{equation}
This factorises as, 
\begin{equation}\label{LanalyticSolution}
\mathcal{L} [f] = \Pi_{j=1}^{m} \int_0^{\infty}
dt_{j} f_{j} (t_{j})e^{-st_{j}} 
\end{equation}
Giving a general analytical solution as,
\begin{equation}\label{analyticSolution}
f(t) = \mathcal{L}^{-1} \left\{ \Pi_{j=1}^{m} 
           \mathcal{L} \left[ f_{j}(t_{j})  \right] \right\}
\end{equation}
where $\mathcal{L}^{-1}$ is the inverse Laplace transform, 
and $\mathcal{L}\left[ f_j(t_j) \right] = \int_0^{\infty} dt_j f_j(t_j) e^{-st_j}$ with the same variable $s$ for each 
value of $j$. 
Eq. \ref{LanalyticSolution} is similar to the relationship 
between moment generating functions $M_i(s)=\sum_{t_i=0}^{\infty} e^{st_i} p_i(t_i)$ 
of discrete probability distributions $p_i(t_i)$, and the moment 
generating function $M(s)$ for $t=\sum_{i=1}^m t_i$, that has, 
\begin{equation}\label{momEq}
M(s) = \Pi_{i=1}^m M_i(s) 
\end{equation}
whose derivation is analogous to Eq. \ref{momEq} but with integrals replaced by sums. 
The survival and hazard functions for $f(t)$ can be obtained from 
Eq. \ref{analyticSolution} in the usual way. 
For example,
\begin{equation}\label{seqSi}
\begin{array}{ll}
S_i(t) &= \int_t^{\infty} f_i(y) dy
\\
&= \int_t^{\infty} \mathcal{L}^{-1} \left\{ \Pi_{j=1}^{m_i} 
\mathcal{L} \left[ f_{ij}(t_{ij})  \right] \right\} dy
\end{array}
\end{equation}
that can be used in combination with Eq. \ref{S}.
A number of valuable results are easy to evaluate 
using Eq. \ref{analyticSolution}, as is illustrated in the next section.  

A useful related result is, 
\begin{equation}\label{n-1sum}
f(t) = 
\mathcal{L}^{-1} \left\{ 
\mathcal{L} \left[ f \left( \sum_{j=1}^{n-1} t_j\right) \right]
\mathcal{L} \left[ f_n(t_n) \right] 
\right\} 
\end{equation}
that can be inferred from Eq. \ref{analyticSolution} with $m=2$, 
\begin{equation}\label{fsum2}
f \left( t = t_1+ t_2 \right) = 
\mathcal{L}^{-1} \left\{ 
\mathcal{L} \left[ f_1(t_1) \right]
\mathcal{L} \left[ f_2(t_2) \right] 
\right\} 
\end{equation}
by replacing $f_1(t_1)$ with $f(\sum_{j=1}^{n-1} t_j )$ and $f_2(t_2)$ with $f_n(t_n)$. 
Eq. \ref{fsum2} can be solved using the convolution theorem for Laplace transforms, 
that gives, 
\begin{equation}\label{convolution}
f \left( t= t_1+t_2 \right) =
\int_0^t f_1(\tau) f_2(t-\tau) d\tau 
\end{equation}
which is sometimes easier to evaluate than two Laplace transforms and their inverse. 
In general, solutions can be presented in terms of multiple convolutions if it is 
preferable to do so. 
Eqs. \ref{n-1sum} and \ref{convolution} are particularly useful for combining a known 
solution for the sum of $(n-1)$ samples such as for cancer initiation, with a differently distributed $n$th sample, such as the waiting time to detect a growing cancer. 
A final remark applies to the sum of random variables whose domain extends from 
$-\infty$ to $\infty$, as opposed to the range $0$ to $\infty$ considered so far.
In that case an analogous calculation using a Fourier transform with respect to $t$ 
in Eq. \ref{EqToTransform} leads to analogous results in terms of Fourier 
transforms, with 
$\mathcal{F} [ f_i(t_i) ] = \int_{-\infty}^{\infty} f_i(t_i)e^{imt_i} dt_i$ 
in place of Laplace transforms, resulting in,
\begin{equation}\label{FT}
f(t) = \mathcal{F}^{-1} \left\{ \Pi_{j=1}^{m} 
           \mathcal{F} \left[ f_{j}(t_{j})  \right] \right\}
\end{equation}
Eq. \ref{FT} is mentioned for completeness, but is not used here. 
A general solution to Eq. \ref{analyticSolution} can be given in terms of definite 
integrals, with, 
\begin{equation}\label{gs1}
\begin{array}{ll}
f(t) &= \mathcal{L}^{-1} \left\{ \Pi_{j=1}^{m} 
\mathcal{L} \left[ f_{j}(t_{j})  \right] \right\}
\\
&=t^{m-1} \int_0^1 dy_1 \dots \int_0^1 dy_{m-1} y_1^0 y_2^1 \dots y_{m-1}^{n_m-1} 
\\
& f_1(ty_1 \dots y_{m-1}) 
  f_2(t(1-y_1)y_2 \dots y_{m-1}) 
  f_3(t(1-y_2)y_3 \dots y_{m-1})
\dots 
\\
& f_{m-1}(t(1-y_{m-2})y_{m-1})
  f_m(t(1-y_{m-1}))
\end{array}
\end{equation} 
This can sometimes be easier to evaluate or approximate than Eq. \ref{analyticSolution}. 
A derivation is given in the Supporting Information (S1 Appendix). 
Eq. \ref{gs1} allows a generalised Schwinger/Feynman  
parameterisation \cite{Weinberg} to be derived. 
Writing $g_j(s)=\mathcal{L} \left[ f_j(t_j) \right]$ and taking the Laplace transform of 
both sides of Eq. \ref{gs1}, gives, 
\begin{equation}
\begin{array}{ll}
\Pi_{j=1}^m g_j(s) &= 
\int_0^1dy_1 \dots \int_0^1 dy_{m-1}  y_1^0 y_2^1 \dots y_{m-1}^{n_m-1} 
\\
&\mathcal{L} \left[
t^{m-1} 
\mathcal{L}^{-1} \left\{ g_1(s) \right\}(ty_1 \dots y_{m-1})
\mathcal{L}^{-1} \left\{ g_2(s) \right\}(t(1-y_1)y_2 \dots y_{m-1})
\dots \right. 
\\
&\left.
\mathcal{L}^{-1} \left\{ g_m(s) \right\}(t(1-y_{m-1}))
\right]
\end{array}
\end{equation}
which includes some well known Schwinger/Feynman parameterisations as special cases. 
This is discussed further in the Supporting Information (S1 Appendix).

\section{Modelling sequential events - examples}\label{examples}

In the following examples we consider the time $t=\sum_{i=1}^m t_i$ for a sequence of 
events, with possibly different distributions $f_i(t_i)$ for the time between the 
$(i-1)$th and $i$th event. 
Some of the results are well-known but not usually presented this way, others are new or 
poorly known. 
We will use the Laplace transforms (and their inverses), of, 
\begin{equation}
\mathcal{L}^{-1} \mathcal{L} [ t^p ] = \mathcal{L}^{-1} [ \Gamma(p+1)/s^{p+1} ] = t^p
\end{equation} 
and, 
\begin{equation} 
\mathcal{L}^{-1} \mathcal{L} [ t^p e^{-\mu t} ] = 
\mathcal{L}^{-1} [ \Gamma(p+1)/(s+\mu)^{p+1} ] = t^p e^{-\mu t}
\end{equation}
{\bf Sums of gamma distributed samples (equal rates):}
Using Eq. \ref{analyticSolution}, the sum of $m$ gamma distributed variables with equal rate 
parameters $\mu$, and $f_i(t_i)=\mu^{p_i} t_i^{p_i-1} e^{-\mu t_i}/\Gamma(p_i)$, are distributed as, 
\begin{equation}\label{sumGamma}
\begin{array}{ll}
f(t)=\mathcal{L}^{-1} \left\{ \Pi_{i=1}^m \mathcal{L} \left[ 
\mu^{p_i} \frac{ t_i^{p_i-1} e^{-\mu t_i} }{\Gamma(p_i)]} \right] \right\} 
&= \mathcal{L}^{-1} \left\{ \Pi_{i=1}^m 
\frac{ \mu^{p_i} }{(s+\mu)^{p_i}} \right\} 
\\
&= \mathcal{L}^{-1} \left\{ 
\frac{\mu^{\sum_{i=1}^m p_i}}{(s+\mu)^{\sum_{i=1}^m p_i} } 
\right\}
\\
&= \mu^{\sum_{i=1}^m p_i} 
\frac{ t^{\sum_{i=1}^m p_i } e^{-\mu t} }{ \Gamma\left(\sum_{i=1}^m p_i \right) }
\end{array}
\end{equation}
For a sum of $m$ exponentially distributed variables with $\{ p_i=1 \}$, this simplifies 
to $f(t)=\mu^m t^{m-1}e^{-\mu t}/\Gamma(m)$, a Gamma distribution. 
{\bf Power law approximations:} For many situations such as most diseases, you are unlikely to 
get any particular disease during your lifetime. 
In those cases the probability of survival over a lifetime is close to $1$, and the probability 
density function $f_i=h_i/S_i$, can be approximated by $f_i\simeq h_i$, that in turn can often be 
approximated by a power of time with $f_i \simeq h_i \simeq \mu_i t_i^{p_i}$. 
Then we have, 
\begin{equation}\label{approxMS}
\begin{array}{ll}
f(t)=\mathcal{L}^{-1} \left\{ \Pi_{i=1}^m \mathcal{L} \left[ \mu_i t_i^{p_i} \right] \right\} 
&= \mathcal{L}^{-1} \left\{ \Pi_{i=1}^m  \frac{\mu_i \Gamma(1+p_i)}{s^{1+p_i}}  \right\} 
\\
&= 
\Pi_{i=1}^m \left( \mu_i \Gamma(1+p_i) \right)
\mathcal{L}^{-1} \left\{ \frac{1}{s^{m+\sum_{i=1}^m p_i} } \right\}
\\
&= \Pi_{i=1}^m \left( \mu_i \Gamma(1+p_i)\right) 
\frac{t^{-1+m+\sum_{i=1}^m p_i } }{ \Gamma(m+\sum_{i=1}^m p_i)}
\end{array}
\end{equation}
{\bf The Armitage-Doll model:} A well known example of this approximation Eq. \ref{approxMS}, is (implicitly) in the original approximate solution 
to the Armitage-Doll multi-stage cancer model. 
Taking a constant hazard at each step, and approximating $f_i \simeq h_i=\mu_i$, 
then Eq. \ref{approxMS} gives,
\begin{equation}
f(t)=\mathcal{L}^{-1} \left\{ \Pi_{i=1}^m \mathcal{L} \left[ \mu_i \right] \right\} = 
\left[ \Pi_{i=1}^m \mu_i \right] \frac{t^{m-1 } }{\Gamma(m)} 
\end{equation}
as was used in the original Armitage-Doll paper. 
Note that an equivalent time-dependence can be produced by a different combination of hazard 
functions with $h_i \sim t_i^{p_i}$ and $\tilde{m}$ steps, provided $m=\tilde{m} + \sum_{i=1}^{\tilde{m}} p_i$. 
For example, if $m=6$, there could be $3$ steps with $p=1$, or $2$ steps with $p=2$, or $3$ steps with 
$p_1=0$, $p_2=1$, and $p_3=2$, or some more complex combination. 
If the full pdfs are modelled at each step as opposed to their polynomial approximation, 
then this flexibility is reduced, as is the case for Moolgavkar's exact solution 
to the Armitage-Doll model that is described next. 
{\bf Moolgavkar's exact solution to the Armitage-Doll model:} 
Moolgavkar's exact solution to the Armitage-Doll model is the solution of, 
\begin{equation}\label{MLT}
f(t)=\mathcal{L}^{-1} \left\{ \Pi_{i=1}^m \mathcal{L} \left[ \mu_i e^{-\mu_i t_i} \right] \right\}=
\mathcal{L}^{-1} \left\{ \Pi_{i=1}^m \frac{\mu_i}{s+\mu_i} \right]
\end{equation}
For example, if $n=3$ then,
\begin{equation}
\mathcal{L}^{-1} \left\{ 
\Pi_{i=1}^3 \mathcal{L} 
\left[ \mu_i e^{-\mu_i t_i} \right] 
\right\}=
\mu_1\mu_2\mu_3 \mathcal{L}^{-1} 
\left\{ 
\frac{1}{(s+\mu_1)} \frac{1}{(s+\mu_2)} \frac{1}{(s+\mu_3)} 
\right\}
\end{equation}
Using partial fractions, we can write,
\begin{equation}
\begin{array}{l}
\frac{1}{s+\mu_1} \frac{1}{s+\mu_2} \frac{1}{s+\mu_3} =
\frac{1}{s+\mu_1} \frac{1}{ (\mu_1-\mu_2)(\mu_1-\mu_3) } + 
\frac{1}{s+\mu_2} \frac{1}{ (\mu_2-\mu_1)(\mu_2-\mu_3) } + 
\frac{1}{s+\mu_3} \frac{1}{ (\mu_3-\mu_1)(\mu_3-\mu_2) }
\end{array}
\end{equation}
Allowing the inverse Laplace transforms to be easily evaluated, giving, 
\begin{equation}\label{3terms}
\begin{array}{c}
f(t)=\mathcal{L}^{-1} \left\{ 
\Pi_{i=1}^3 \mathcal{L} 
\left[ \mu_i e^{-\mu_i t} \right] 
\right\}
\\=
\mu_1 \mu_2 \mu_3 \left[
\frac{e^{-\mu_1 t}}{(\mu_1-\mu_2)(\mu_1-\mu_3)} + 
\frac{e^{-\mu_2 t}}{(\mu_2-\mu_1)(\mu_2-\mu_3)} + 
\frac{e^{-\mu_3 t}}{(\mu_3-\mu_1)(\mu_3-\mu_2)} 
	\right] 
\end{array}	
\end{equation} 
Note that the result is independent of the order of sequential events, but unlike 
the approximate solution to the Armitage Doll model \cite{DollArmitage}, the exact solution 
allows less variability in the underlying models that can produce it. 
Also note that the leading order terms of an expansion in $t$ cancel exactly, to give 
identical leading-order behaviour as for a power-law approximation (with $p=0$)
A general solution can be formed using a Schwinger/Feynman 
parameterisation \cite{Weinberg} of, 
\begin{equation}\label{SFparam}
\begin{array}{c}
\Pi_{i=1}^m \frac{1}{\mu_i} = 
\Gamma(m) \int_0^1 dy_1 \int_0^{y_1} dy_2 \dots \int_0^{y_{m-2}} dy_{m-1} 
\frac{1}{(\mu_1 y_{m-1} + \mu_2(y_{m-2}-y_{m-1}) + \dots + \mu_m(1- y_1) )^m}
\end{array}
\end{equation}
Replacing $\mu_i$ with $s+\mu_i$ in Eq. \ref{SFparam}, then we can write Eq. 
\ref{MLT} as, 
\begin{equation}
\begin{array}{l}
\mathcal{L}^{-1} \left\{ \Pi_{i=1}^m \frac{\mu_i}{s+\mu_i} \right\} 
\\
= \left( \Pi_{i=1}^m \mu_i \right) \Gamma(m) \times 
\\
\int_0^1 dy_1 \int_0^{y_1} dy_2 \dots \int_0^{y_{m-2}} dy_{m-1}
\mathcal{L}^{-1} \left\{ 
\frac{1}{(s + \mu_1 y_{m-1} + \mu_2(y_{m-2}-y_{m-1}) + \dots + \mu_m(1- y_1) )^m}
\right\}
\\
= \left( \Pi_{i=1}^m \mu_i \right) t^{m-1} \times
\\
\int_0^1 dy_1 \int_0^{y_1} dy_2 \dots \int_0^{y_{m-2}} dy_{m-1}
e^{ - ( \mu_1 y_{m-1} + \mu_2(y_{m-2}-y_{m-1}) + \dots + \mu_m(1- y_1) )t }
\end{array}
\end{equation}
(which is simpler, but equivalent in effect, to repeatedly using the convolution formula). 
Completing the integrals will generate Moolgavkar's solution for a given value of $m$. 
For example, taking $m=3$ and integrating once gives, 
\begin{equation}
\mathcal{L}^{-1} \left\{ \Pi_{i=1}^3 \frac{\mu_i}{s+\mu_i} \right\} 
= \frac{te^{-\mu_3 t}}{(\mu_2-\mu_1)} \int_0^1 dx_1 
\left( e^{-x_1t(\mu_1-\mu_3)} - e^{-x_1 t(\mu_2-\mu_3)} \right) 
\end{equation}
Integrating a second time, and simplifying, gives Eq. \ref{3terms}. 
The relationships between Schwinger/Feynman parameterisations, Laplace transforms, and 
the convolution theorem are discussed further in the Supplementary Information (S1 Appendix).
Moolgavkar \cite{Moolgavkar} used induction to provide an explicit formula for $f(t)$, with, 
\begin{equation}\label{MoolgavkarSoln}
f(t) = \left( \Pi_{i=1}^m \mu_i \right) \sum_{i=1}^m \chi_i(m) e^{-\mu_i t} 
\end{equation}
where, 
\begin{equation}\label{chi}
\chi_{i}(m) = \frac{1}{(\mu_1-\mu_i)(\mu_2-\mu_i) \dots 
(\mu_{i-1}-\mu_i)(\mu_{i+1}-\mu_i) \dots (\mu_m-\mu_i) }
\end{equation}
For small times the terms in a Taylor expansion of Eq. \ref{MoolgavkarSoln} cancel 
exactly, so that $f(t) \simeq \left( \Pi_{i=1}^m \mu_i \right) t^{m-1}$, as expected. 
This feature could be useful for approximating a normalised function when the 
early-time behaviour approximates an integer power of time.
Further uses of Moolgavkar's solution are discussed next. 
{\bf Sums of gamma distributed samples (with different rates):}
A useful mathematical result can be found by combining the Laplace transform of 
Moolgavkar's solution Eq. \ref{MoolgavkarSoln} 
for $f\left( t= \sum_{i=1}^m t_i \right)$ 
with Eq. \ref{MLT}, to give an explicit formula for a partial fraction decomposition 
of the product $\Pi_{i=1}^m \frac{1}{s+\mu_i}$, as, 
\begin{equation}\label{parfrac}
\Pi_{i=1}^m \frac{1}{s+\mu_i} = 
\sum_{i=1}^m \frac{\chi_{i}(m)}{s+\mu_i} 
\end{equation}
This can be useful in various contexts.  
For example, 
consider $m$ Gamma distributions $f_i(t_i)=\mu_i^{p_i}t_i^{p_i-1}e^{-\mu_i t_i}/\Gamma(p_i)$ 
with different integer-valued shape parameters $p_i$, and 
$\mathcal{L} [f_i]=\mu_i^{p_i}/(s+\mu_i)^{p_i}$. 
Eq. \ref{analyticSolution} gives 
$f(t)= \left( \Pi_{i=1}^m \mu_i^{p_i} \right) \mathcal{L}^{-1} 
\left\{ \Pi_{i=1}^m 1/(s+\mu_i)^{p_i} \right\}$, 
so firstly use the integer-valued property of $\{ p_i \}$ to write, 
\begin{equation}
\begin{array}{ll}
\mathcal{L}^{-1} \left\{ 
\Pi_{i=1}^m \frac{1}{(s+\mu_i)^{p_i}} 
\right\} 
&= \mathcal{L}^{-1} \left\{ 
\Pi_{i=1}^m 
\frac{(-1)^{p_i-1}}{(p_i-1)!} \frac{\partial^{p_i-1}}{\partial \mu_i^{p_i-1}} 
\frac{1}{(s+\mu_i)} 
\right\}
\\
&= \mathcal{L}^{-1} \left\{ 
\Pi_{j=1}^m 
\frac{(-1)^{p_j-1}}{(p_j-1)!} \frac{\partial^{p_j-1}}{\partial \mu_j^{p_j-1}} 
\Pi_{i=1}^m
\frac{1}{(s+\mu_i)} 
\right\}
\end{array}
\end{equation}
where the product of differential operators can be taken outside the product of Laplace
transforms because $\partial/\partial \mu_i (1/(s+\mu_j))$ is zero for $i\neq j$. 
Using Eq. \ref{parfrac} we can replace the product of Laplace transforms with a sum, giving,  
\begin{equation}
\mathcal{L}^{-1} \left\{ 
\Pi_{i=1}^m \frac{1}{(s+\mu_i)^{p_i}} 
\right\} 
=
\mathcal{L}^{-1} \left\{ 
\Pi_{j=1}^m 
\frac{(-1)^{p_j-1}}{(p_j-1)!}
\frac{\partial^{p_j-1}}{\partial \mu_j^{p_j-1}} 
\sum_{i=1}^{m}
\frac{\chi_i(m)}{(s+\mu_i)} 
\right\}
\end{equation}
The Laplace transform has now been simplified to a sum of 
terms in $1/(s+\mu_i)$, whose inverse Laplace transforms are easy 
to evaluate. 
Taking the inverse Laplace transform 
$\mathcal{L}^{-1}[1/(s+\mu_i)]=e^{-\mu_i t}$, and including the product $\Pi_{i=1}^m \mu_i^{p_i}$, gives,
\begin{equation}\label{ivalGam}
f(t) = \left( \Pi_{i=1}^m \mu_i^{p_i}  \right) 
 \Pi_{j=1}^m \frac{(-1)^{p_j-1}}{(p_j-1)!} \frac{\partial^{p_j-1}}{\partial \mu_j^{p_j-1}} 
 \sum_{i=1}^{m} \chi_i(m) e^{-\mu_i t} 
\end{equation}
as a general solution for sums of Gamma distributed samples with integer-valued 
shape parameters $p_i$ (and arbitrary rate parameters $\mu_i$). 
Eq. \ref{ivalGam} is most easily evaluated with a symbolic algebra package. 
If $p_i=p$ are equal, then Eq. \ref{ivalGam} may be simplified further by writing, 
\begin{equation}
f(t) = \left( \Pi_{i=1}^m \mu_i^{p}  \right) 
\sum_{i=1}^{m} 
 \frac{(-1)^{p-1}}{(p-1)!} \frac{\partial^{p-1}}{\partial \mu_i^{p-1}}
\Pi_{j\neq i} \frac{(-1)^{p-1}}{(p-1)!} \frac{\partial^{p-1}}{\partial \mu_j^{p-1}} 
\left[
\chi_i(m) e^{-\mu_i t}
\right]
\end{equation}
and noting that, 
\begin{equation}
\Pi_{j\neq i} \frac{(-1)^{p-1}}{(p-1)!} \frac{\partial^{p-1}}{\partial \mu_j^{p-1}} 
\left[ \chi_i(m) e^{-\mu_i t} \right] = \chi_i(m)^{p} e^{-\mu_i t} 
\end{equation}
because for $j\neq i$ there is exactly one factor $1/(\mu_j-\mu_i)$ in $\chi_i(m)$. 
This leaves, 
\begin{equation}\label{sameP}
f(t) = \left( \Pi_{i=1}^m \mu_i^{p}  \right) 
\sum_{i=1}^{m} 
\frac{(-1)^{p-1}}{(p-1)!} \frac{\partial^{p-1}}{\partial \mu_i^{p-1}}
\left[
\chi_i(m)^{p} e^{-\mu_i t} 
\right]
\end{equation}
for sums of Gamma distributed samples with the same integer-valued 
shape parameter $p$ (and arbitrary rate parameters $\mu_i$). 
For example, if $p=1$ then Eq. \ref{sameP} becomes Moolgavkar's Eq. \ref{MoolgavkarSoln}. 
Alternatively, if e.g. $p=2$, then we have, 
\begin{equation}\label{ivalGam1} 
f(t) = 
\left( \Pi_{i=1}^m \mu_i^2 \right) 
\sum_{i=1}^m  
\chi_i(m)^2 e^{-\mu_i t} \left[ t - 2 \sum_{j \neq i} \frac{1}{(\mu_j-\mu_i)} \right]  
\end{equation}
for the sum of Gamma distributions with shape parameters $p=2$ and arbitrary rate parameters, and $\chi_i(m)$ as defined in Eq. \ref{chi}. 
If we also let e.g. $m=2$, $\mu_2=\mu_1+\epsilon$, and  
$\epsilon \rightarrow 0$, then Eq. \ref{ivalGam1} tends to 
$\mu_1^4 t^3 e^{-\mu_1 t}/3!$, for the sum of two 
Gamma distributed  variables with rate $\mu_1$ and $p=2$, 
in agreement with Eq. \ref{sumGamma}. 
{\bf Sums of samples with different distributions:} 
An advantage of the method described above, is that it is often easy to calculate pdfs for sums of differently distributed samples.
For the first example, consider two samples from the same (or very similar) exponential distribution, and a third from a different exponential distribution.
The result can be obtained by writing $\mu_3=\mu_2+\epsilon$ in Eq. \ref{3terms}, and letting 
$\epsilon \rightarrow 0$. 
Considering the terms involving exponents of $\mu_2$ and $\mu_3$, 
\begin{equation} 
\begin{array}{ll}
\frac{ e^{-\mu_2 t} }{ (\mu_2-\mu_1)(\mu_2-\mu_3) } + 
\frac{ e^{-\mu_3 t} }{ (\mu_3-\mu_1)(\mu_3-\mu_2) } 
&= \frac{ e^{-\mu_2 t} }{(\mu_2-\mu_1)\epsilon} \left( 
	- 1 
	+ \frac{ e^{-\epsilon t} }{ 1+\epsilon/(\mu_2-\mu_1) } 
	\right)
\\
&= \frac{ e^{-\mu_2 t} }{ \mu_2-\mu_1 } 
\left( -1 + \left( 1- \epsilon t - \frac{\epsilon}{\mu_2-\mu_1} + O\left( \epsilon^2 \right)  \right) \right)
\\
&= \left[  - \frac{te^{-\mu_2 t}}{\mu_2-\mu_1} - \frac{e^{-\mu_2 t}}{(\mu_2-\mu_1)^2}\right] \left( 1 + O\left( \epsilon \right)  \right)
\end{array}
\end{equation}
Using Eq. \ref{3terms} and letting $\epsilon \rightarrow 0$, gives, 
\begin{equation}\label{Eq2exp}
\begin{array}{c}
\mu_1 \mu_2 \mu_3 \left[
\frac{e^{-\mu_1 t}}{(\mu_1-\mu_2)(\mu_1-\mu_3)} + 
\frac{e^{-\mu_2 t}}{(\mu_2-\mu_1)(\mu_2-\mu_3)} + 
\frac{e^{-\mu_3 t}}{(\mu_3-\mu_1)(\mu_3-\mu_2)} 
\right] 
\\
\rightarrow \mu_1 {\mu_2}^2 \left[
\frac{e^{-\mu_1 t} - e^{-\mu_2 t} }{(\mu_1-\mu_2)^2} +  
\frac{t e^{-\mu_2 t}}{(\mu_1-\mu_2)} 
\right] 
\end{array}
\end{equation}
for the sum of three exponentially distributed variables, when exactly two have the same rate.  
Taking $\mu_2=\mu_1+\epsilon$ and letting $\epsilon \rightarrow 0$ in Eq. \ref{Eq2exp}, gives 
a Gamma distribution $\mu_1^3 t^2 e^{-\mu_1 t}/2$, as it should for the sum of three exponentially 
distributed variables with equal rates (see Eq. \ref{sumGamma} with $\{p_i=0\}$). 
More generally, it can be seen that a sum of exponentially distributed samples with different rates, smoothly approximate a gamma distribution as the rates become increasingly similar, 
as expected from Eq. \ref{sumGamma}. 

\begin{figure}[ht]
  \centering{\includegraphics[width=0.5\linewidth]{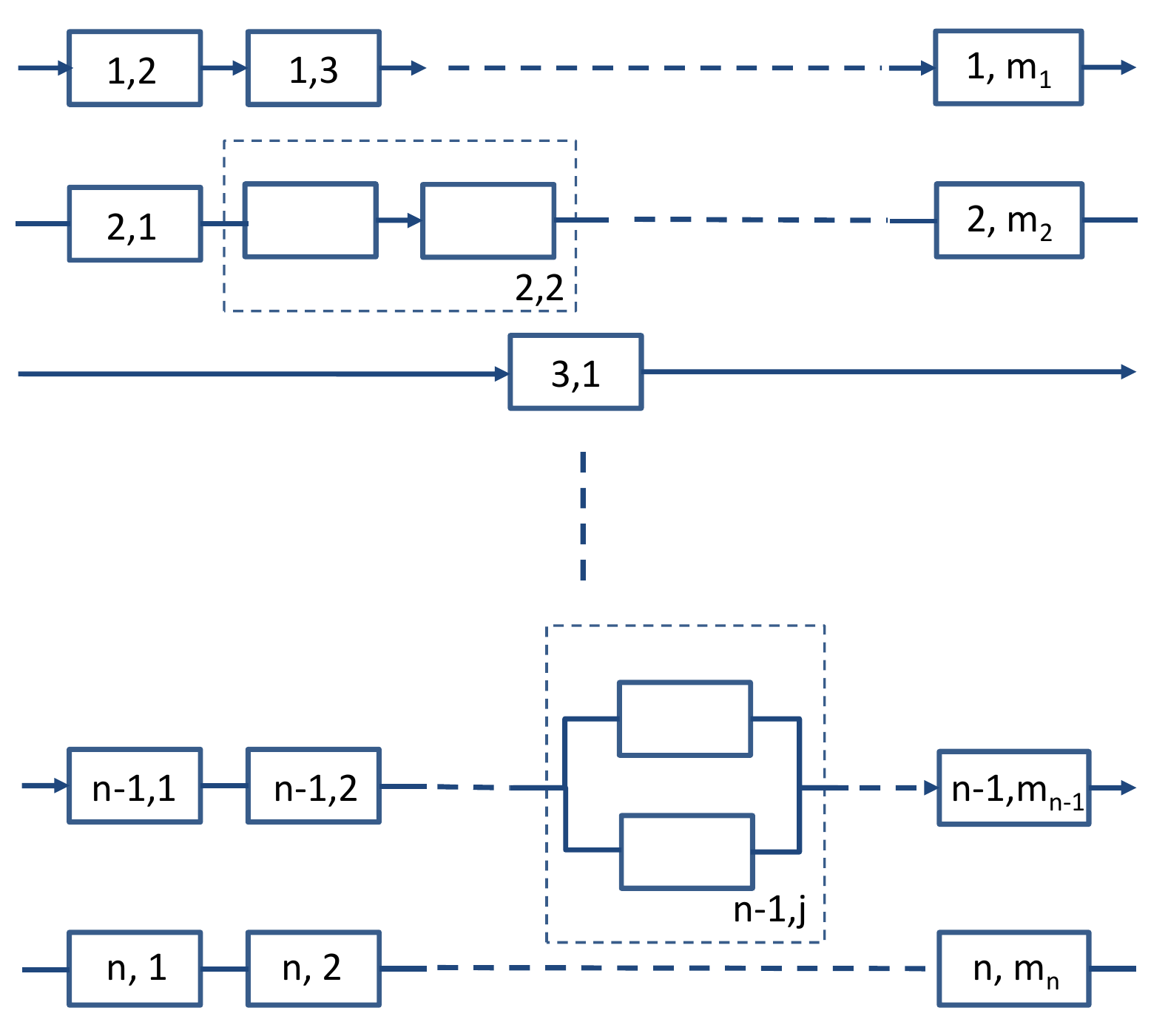}}
  \vspace{0.4cm}
  \caption{\sl Overall failure risk can be modelled as sequential steps (e.g. ($1$,$1$) to 
  ($1$,$m_1$) 
  using Eq. \ref{S_i_indp}), and non-sequential steps (e.g. ($n$,$1$) to ($n$,$m_n$) 
  using Eq. \ref{analyticSolution}), that may be dependent on each other 
  (e.g. Eq. \ref{fABC}). For the purposes of modelling, a sequence of dependent 
  or multiple routes can be regarded as a single step (e.g. ($2$,$2$) or ($n-1$,$j$)).}
  \label{fig4}
\end{figure}

{\bf Failure involving a combination of sequential and non-sequential steps:}
If a path to failure involves a combination of sequential and non-sequential steps, then 
the necessary set of sequential steps can be considered as one of the non-sequential steps, 
with overall survival given by Eq. \ref{S} and the survival for any sequential set of 
steps calculated from Eq. \ref{seqSi} (Fig \ref{fig4}).  

\section{Clonal-expansion cancer models}

Clonal expansion is thought to be an essential element of cancer progression 
\cite{Greaves2012}, and can modify the timing of cancer onset and detection 
\cite{LuebeckMoolgavkar2002, Michor2006, MezaLuebeck2008, Beerenwinkel2007, Beerenwinkel2015, Altrock2015}. 
The growing number of cells at risk increases the probability of the next step in a sequence of mutations occurring, and if already cancerous, then it increases the likelihood of detection.  
Some cancer models have a clonal expansion of cells as a rate-limiting step 
\cite{LuebeckMoolgavkar2002, Michor2006, MezaLuebeck2008}.  
For example, Michor et al. \cite{Michor2006} modelled clonal expansion of myeloid leukemia 
as logistic growth, with the likelihood of cancer detection (the hazard function), being 
proportional to the number of cancer cells. 
This gives a survival function for cancer detection of, 
\begin{equation}
S_i(t) = \exp \left( - a \int_0^t x(y) dy \right)
\end{equation}
where, 
\begin{equation}
x(t) = \frac{1}{1+(N-1)e^{-ct}}
\end{equation}
$a$, $c$, are rate constants, and $N$ is the total number of cells prior to cancer 
initiation. 
Noting that $\int_0^t x(y) dy = \log ( e^{ct} + (N-1) )/c \rightarrow t $, 
as $t \rightarrow \infty$ and $x(t) \rightarrow 1$, then the tail of the 
survival curve falls exponentially towards zero with time. 
Alternatively, we might expect the likelihood of cancer being diagnosed to continue to 
increase with time since the cancer is initiated. 
For example, a hazard function that is linear in time would give a Weibull distribution with 
$S(t) = e^{-at^2}$. 
It is unlikely that either this or the logistic model would be an equally good 
description for the detection of all cancers, although they may both be an improvement 
on a model without either. 
Both models have a single peak, but differ in the tail of their distribution, that falls as 
$\sim e^{-act}$ for the logistic model and $\sim e^{-at^2}$ for the Weibull model. 
Qualitatively, we might expect a delay between cancer initiation and the possibility of 
diagnosis, and diagnosis to occur almost inevitably within a reasonable time-period. 
Therefore a Weibull or Gamma distributed time to diagnosis may be reasonable 
for many cancers, with the shorter tail of the Weibull distribution 
making it more suitable approximation for cancers whose diagnosis is almost inevitable. 
(The possibility of misdiagnosis or death by another cause is not considered here.)
For example, noting that Moolgavkar's solution is a linear combination of exponential distributions, to combine it with a Weibull distribution for cancer detection 
$f_1(t_1)=-d/dt_1 ( e^{-bt_1^2/2} )$, 
we can consider a single exponential term at a time. 
Taking $f_2(t_2)=ae^{-at_2}$, and using the convolution formula Eq. \ref{convolution}, 
we get,  
\begin{equation}
\begin{array}{ll}
f(t=t_1+t_2) 
&= \mathcal{L}^{-1} \left\{ L[f_1(t_1)] L[f_2(t_2)] \right\}
\\
&= a \int_0^t e^{-a(t-y)} \left( -\frac{d}{dy} e^{-by^2/2} \right) dy
\\
&= a \left( e^{-at} - e^{-bt^2/2} \right)
+ a^2 e^{-at} e^{a^2/2b} \int_0^t e^{-\frac{b}{2} \left( y-\frac{a}{b} \right)^2 } dy
\end{array}
\end{equation}
where we integrated by parts to get the last line. 
This may be written as, 
\begin{equation}
\begin{array}{ll}
f(t) &= a \left( e^{-at} - e^{-bt^2/2} \right)
\\
&+ a^2 e^{-at} e^{a^2/2b} \sqrt{ \frac{\pi}{2b} } \mbox{ }
\mbox{erf} \left( \sqrt{\frac{b}{2}} \frac{a}{b} \right) 
\\
&+ a^2 e^{-at} e^{a^2/2b} \sqrt{ \frac{\pi}{2b} } \left\{
\begin{array}{l}
- \mbox{erf} \left( \sqrt{\frac{b}{2}} \left( \frac{a}{b} - t \right) \right) 
\mbox{  } t < \frac{a}{b}
\\
+ \mbox{erf} \left( \sqrt{\frac{b}{2}} \left( t - \frac{a}{b} \right) \right) 
\mbox{  } t \geq \frac{a}{b} 
\end{array}
\right.
\end{array}
\end{equation}
with $\mbox{erf}(x)=\frac{2}{\sqrt{\pi}} \int_0^x e^{-z^2} dz$. 
Similarly for a Gamma distribution with $f_1=b^p t^{p-1} e^{-bt}/\Gamma(p)$ and 
an exponential, $f_2(t_2)=ae^{-at}$, then assuming $b>a$, 
\begin{equation}
\begin{array}{ll}
f(t) 
&= \frac{b^p a}{\Gamma(p)} \int_0^t y^{p-1} e^{-by} e^{-a(t-y)} dy 
\\
&= b^p a \frac{e^{-at}}{(b-a)^p} \frac{1}{\Gamma(p)} \int_0^{t(b-a)} u^{p-1} e^{-u} du 
\\
&= b^p a \frac{e^{-at}}{(b-a)^p} \gamma(p,t(b-a))
\end{array}
\end{equation}
where $\gamma(p,t(b-a))$ is the normalised lower incomplete Gamma function, which is 
available in most computational mathematics and statistics packages. 
If $a>b$ then $f_1$ and $f_2$ must be exchanged and the result is most easily evaluated 
numerically. 

\section{Cascading failures with dependent sequences of events}\label{cascading}

Now consider non-independent failures, where 
the failure of A changes the probability of a failure in B or C. 
In general, if the paths to failure are not independent of each other then the situation cannot be described by Eq. \ref{S}. 
Benjamin Cairns suggested exploring the following example - 
if step 1 of A prevents step 1 of B and vice-versa, then only one path can be followed. 
If the first step occurs at time $t_1$, the pdf for failure at time $t$ is: $f(t)=S_A(t_1)f_B(t)+S_B(t_1)f_A(t)$, where $f_A(t)$ and $f_B(t)$ 
are the pdfs for path A and B if they were independent. 
This differs from Eq. \ref{S} that has, $f(t)=-dS/dt=S_A(t)f_B(t)+S_B(t)f_A(t)$, 
that is independent of $t_1$.  
As a consequence, Eq. \ref{S} may be inappropriate to describe phenomenon such as  
survival in the presence of natural selection, where competition for the same resource  
means that not all can survive.
In some cases it may be possible to include a different 
model for the step or steps where Eq. \ref{S} fails, analogously to the clonal expansion model \cite{Michor2006} described in Section \ref{examples}. 
But in principle, an alternative model may be required. 
We will return to this point in Section \ref{evolution}. 
The rest of this section limits the discussion to situations where the paths to failure are independent, but where the 
failure-rate depends on the order of events. 
Important humanitarian examples are ``cascading hazards'' \cite{AghaKouchak}, where the risk of a disaster such as a mud slide is 
vastly increased if e.g. a wildfire occurs before it. 
An equivalent scenario would require $m$ parts to fail for the system to fail, 
but the order in which the parts fail, modifies the probability of subsequent component failures. 
As an example, if three components A, B, and C, must fail, then we need to 
evaluate the probability of each of 
the 6 possible routes in turn, and obtain the overall failure probability from Eq. \ref{S}. 
Assuming the paths to failure are independent, then there are $m!$ routes, giving $6$ in this example.  
Writing the 6 routes as, 
$1$=ABC, 
$2$=ACB,
$3$=BAC, 
$4$=BCA,
$5$=CAB, 
$6$=CBA, and reading e.g. ABC as ``A, then B, then C'', the survival probability is,
\begin{equation}\label{SABC}
S(t) = \Pi_{i=1}^6 S_{i}(t) 
\end{equation}
For failure by a particular route $ABC$ we need the probability 
for the sequence of events, 
$A \mbox{\&} (\overline{B \mbox{\&} C})$, 
then $(B \mbox{\&} \bar{C})|A$, 
then $C| (AB)$. 
We can calculate this using Eq. \ref{analyticSolution}, for example giving, 
\begin{equation}\label{fABC}
f_{ABC}(t) = 
\mathcal{L}^{-1} \left\{
\mathcal{L} \left[ f_{A\&(\overline{B\&C})} (t_1) \right]
\mathcal{L} \left[ f_{(B\&\bar{C}) | A} (t_2) \right]
\mathcal{L} \left[ f_{C | (AB)} (t_3) \right]
\right\}
\end{equation}
from which we can construct $S_{1}(t) = \int_t^{\infty} f_{ABC}(y) dy$. 
Although in principle every term in e.g. Eqs. \ref{SABC} and \ref{fABC} need 
evaluating, there will be situations where results simplify. 
For example, if one route is much more probable than another - e.g. if it is approximately 
true that landslides only occur after deforestation, that may be due to fire, then we only 
need to evaluate the probability distribution for that route. 
As another example, if all the $f_i$ are exponentially distributed with different rates, 
then $f_{ABC}$ will be described by Moolgavkar's solution. 
A more striking example is when there are very many potential routes to failure, as for 
the Armitage-Doll model where there are numerous stem cells that can cause cancer. 
In those cases, if the overall failure rate remains low, then the $f_i(t)$ in Eq. \ref{fABC}  
must all be small with $S \simeq 1$ and $f \simeq h$, and can often be approximated by 
power laws. 
For that situation we have a general result that $f_{i}$, $F_{i}$, and $H_i$ 
will be a powers of time, and Eq. \ref{SH} gives, 
\begin{equation}
S(t) \simeq \exp \left\{ - \sum_{i=1}^n a_i t^{p_i} \right\} 
\end{equation}
for some $a_i>0$ and $p_i>0$. 
Then $F(t)=1-S(t)$, $f(t)=-dS/dt$, and $h(t)\simeq f(t)$, can be approximated by 
a sum of power series in time. 
If one route is much more likely than the others then both $f(t)$ and $h(t)$ can be 
approximated as a single power of time, with the approximation best at early times, 
and a cross-over to different power-law behaviour at later times.  

\section{Cancer evolution, the tissue micro-environment, and model limitations}\label{evolution}

Cancer is increasingly viewed as an evolutionary process that is influenced by 
a combination of random and carcinogen-driven genetic and epigenetic changes 
\cite{DollArmitage, Peto2016, Nowell1976, Cairns1975, Maley2006, Gatenby2009, Greaves2012, Tomasetti2015, Tomasetti2017}, 
and an evolving tissue micro-environment 
\cite{Hines2011,Rozhok2015,Maley2017,Gatenby2017}. 
Although there is evidence that the number of stem cell divisions is more important for 
cancer risk than number of mutations \cite{L-LJan2018, L-LMar2018}, 
the recognition that cells in a typical cancer are  
functionally and genetically diverse has helped explain cancers' resistance to treatment, 
and is suggesting  
alternative strategies to tackle the disease through either adaptive therapies 
\cite{GatenbyAdapt2009,Silva2012, Enrique-Navas2016,Zhang2017} 
or by modifying the tissue's  micro-environment 
\cite{GatenbyGillies2008,Bissel2011,Rozhok2015,Gatenby2017}.
This highlights two limitations of the multi-stage model described here.
{\bf Evolution:} 
As noted in Section \ref{cascading}, Eq. \ref{S} cannot necessarily model a competitive 
process such as natural selection, where the growth of one cancer variant can inhibit 
the growth of another.
If the process can be described through a series of rate-limiting steps, then we could still 
approximate it with a form of Eq. \ref{analyticSolution}. 
Otherwise, the time-dependence of a step with competitive evolutionary processes may need to be 
modelled differently \cite{Altrock2015, Beerenwinkel2015}, such as with a Wright-Fisher model \cite{Beerenwinkel2007, Beerenwinkel2015}, or with an approximation such 
as the logistic model used to describe myeloid leukemia \cite{Michor2006}. 
As emphasised by some authors 
\cite{Horvath2013,Rozhok2015}, a large proportion of 
genetic alterations occur before adulthood. 
Therefore it seems possible that some routes to cancer could be determined prior to 
adulthood, with genetic mutations and epigenetic changes 
in childhood either favouring or inhibiting the possible 
paths by which adult cancers could arise.
If this led to a given cancer type occurring with a small number of sufficiently different 
incident rates, then it might be observable in a population's incidence data as a mixture of 
distributions. 
{\bf Changing micro-environment:}    
Another potential limitation of the model described in Section \ref{order} is that the time to failure at 
each step is independent of the other failure times, and of the time at which that step becomes at risk. 
If the tissue micro-environment is changing with time, then this assumption fails, and the 
failure rate at each step is dependent on the present time. 
This prevents the factorisation of the Laplace transform used in Eqs. 
\ref{EqToTransform}-\ref{LanalyticSolution}, that led to 
Eq. \ref{analyticSolution} for failure via $m$ sequential steps.    
We can explore the influence of a changing micro-environment with a perturbative approximation.
The simplest example is to allow the $\{ \mu_j \}$ in the Armitage-Doll model to depend linearly 
on the time $\sum_{k=1}^j t_k$ at which step $j$ is at risk. 
Then the Armitage-Doll approximation of $f_j(t_j) \simeq \mu_j$ for $\mu_j t_j \ll 1$, is 
replaced by 
\begin{equation}
f_j(t_j|t_{j-1}, \dots , t_1 ) \simeq \mu_{j0} + \mu_{j1} \sum_{k=1}^j t_k 
\end{equation}
The calculation in Section \ref{order} is modified, with, 
\begin{equation}
P(T_1=t_1, \dots , T_m=t_m)
= 
f_m( t_m | t_{m-1}, \dots , t_1 ) \dots f_2(t_2|t_1) f_1(t_1) 
\end{equation}
giving, 
\begin{equation}\label{AD2}
\begin{array}{ll}
P(T_1=t_1, \dots , T_m=t_m) 
&= \Pi_{j=1}^m \left( \mu_{j0} + \mu_{j1}\sum_{k=1}^j t_k \right) 
\\
&= a_0 + \sum_{j=1}^{m} a_j t_j^{m-j+1}
\end{array}
\end{equation}
with $a_0=\Pi_{j=1}^m \mu_{j0}$, and 
$\{a_j\}$ being sums of products of $j-1$ factors from $\{\mu_{j0}\}$ and $m-j+1$ factors 
from $\{\mu_{k1}\}$.
Replacing $\Pi_{j=1}^m f_j(t_j)$ in Eqs. \ref{EqToTransform} and \ref{LEqToTransform}, 
with the right-side of Eq. \ref{AD2}, and evaluating the $m$ integrals then gives, 
\begin{equation}
\mathcal{L} \left[ f \right] = 
\frac{a_0}{s^m} + 
\sum_{j=1}^m a_j \frac{\Gamma(m-j+2)}{s^{m-j+2}} \frac{1}{s^{m-1}}
\end{equation}
with solution, 
\begin{equation}\label{fmen}
f(t) = a_0 \frac{t^{m-1}}{\Gamma(m)} + 
\sum_{j=1}^m a_j \frac{\Gamma(m-j+2)}{\Gamma(2m-j+1)}t^{2m-j} 
\end{equation}
If the tissue micro-environment is changing rapidly enough that a term $a_j t_j^{2m-j}$ becomes 
comparable to or larger than $a_0 t^{m-1}$, then the solution to Eq. \ref{fmen} can behave like a 
larger power of time than the usual $m-1$ for $m$ rate-limiting steps. 
It is even possible for the incidence rate to slow or even {\sl decrease}, if 
coefficients in Eq. \ref{fmen} are negative. 
The example illustrates that if the micro-environment modifies cancer risk  
and is changing over a person's lifetime, then it 
has the potential to strongly influence 
the observed rate of cancer incidence. 
The argument can be repeated with less generality or greater sophistication, 
e.g. expanding the coefficients $\mu_i$ in the terms $\exp(-\mu_j t_j)$ that appear 
in Moolgavkar's model. 
Such models will have a complex relationship between their coefficients that might make them 
identifiable from cancer incidence data. 
This goes beyond the intended scope of this paper.

\section{Conclusions}

The purpose of this article is to provide a simple mathematical framework to 
describe existing multi-stage cancer models, that is easily adaptable to model 
events such as failure of complex systems, cascading disasters, and 
the onset of disease. 
The key formulae are Eqs. \ref{S}, \ref{S_i}, and 
\ref{analyticSolution} or equivalently \ref{seqSi}, and a  
selection of analytical results are given to illustrate their use.
Limitations of the multi-stage model are discussed in Sections \ref{cascading} and 
\ref{evolution}. 
The examples in Section \ref{examples} can be combined in numerous ways to construct 
a wide range of models. 
Together the formulae are intended to provide a comprehensive toolkit for developing 
conceptual and quantitative models to describe 
failure, disaster, and disease. 

\section*{Acknowledgments}
Thanks to Benjamin Cairns and Andrii Rozhok for helpful comments. 
This research was funded by a fellowship from the Nuffield Department of Population 
Health, University of Oxford, and with support from Cancer Research UK (grant no. C570/A16491). 

\section*{Supporting Information}

\subsection*{S1 Appendix - Derivation of Eq. 23, and its relationship to Schwinger/Feynman parameterisations}\label{appendix1}

\noindent
The solution of Eq. 16 can be written in terms of multiple definite 
integrals that are sometimes easier to evaluate or approximate than directly evaluating 
Eq. 16. 
It is equivalent to expressing the solution as multiple convolutions using 
Eq. 21, and changing variables appropriately.  
The equation is obtained by Taylor expanding all functions before taking their 
Laplace transform, inverting the Laplace transform of the product of all terms 
(which is easy to do for the powers of time that appear in a Taylor expansion), 
then using a product of Beta functions to factorise and re-sum the resulting 
expression. 
In mathematical notation, 
\begin{equation}\label{app1}
\begin{array}{ll}
\mathcal{L}^{-1} 
\left\{ 
\Pi_{j=1}^{m} 
\mathcal{L} \left[ f_{j}(t_{j})  \right] 
\right\} 
&= 
\mathcal{L}^{-1} 
\left\{ 
\Pi_{j=1}^{m} 
\mathcal{L} \left[ \sum_{n_j=0}^{\infty} f_{j,n_j} \frac{t^{n_j}}{n_j!}  \right] 
\right\} 
\\
&= 
\mathcal{L}^{-1} 
\left\{ 
\sum_{n_1=0}^{\infty} \dots \sum_{n_m=0}^{\infty} 
\frac{f_{1,n_1}}{s^{n_1+1}} \dots  \frac{f_{m,n_m}}{s^{n_m+1}}  
\right\}      
\\
&= 
\sum_{n_1=0}^{\infty} \dots \sum_{n_m=0}^{\infty} 
f_{1,n_1} \dots f_{m,n_m} \frac{t^{-1+\sum_{i=1}^m(n_i+1)}}{\Gamma( \sum_{i=1}^m (n_i+1))}  
\end{array}
\end{equation}
where $f_{i,n_j} = \partial^{n_j} f_i(t_i) / \partial t^{n_j} |_{t_i=0}$. 
Now noting that the Beta function has, 
\begin{equation}
\int_0^1 u^{m-1}(1-u)^{n-1} du = \frac{\Gamma(m)\Gamma(n)}{\Gamma(m+n)} 
\end{equation}
we can write, 
\begin{equation}
\frac{1}{\Gamma\left( \sum_{i=1}^m (n_i + 1) \right)}
= 
\frac{1}{\Gamma\left( \sum_{i=1}^{m-1} (n_i + 1) \right)}
\frac{1}{\Gamma\left( n_m + 1 \right)}
\int_0^1 y_{m-1}^{-1+\sum_{i=1}^{m-1} (n_i+1)} (1-y_{m-1})^{n_{m}} 
\end{equation}
Repeatedly using this gives, 
\begin{equation}\label{app2}
\begin{array}{ll}
\frac{1}{\Gamma\left( \sum_{i=1}^m (n_i + 1) \right)}
&= 
\left( \Pi_{i=1}^m \frac{1}{\Gamma(n_i+1)} \right)
\int_0^1 dy_1 \dots \int_0^1 dy_{m-1} y_1^0 y_2^1 y_3^2 \dots y_{m-1}^{m-2} \times 
\\
& (1-y_1)^{n_2} \dots 
(1-y_{m-1})^{n_m} 
\times 
(y_1 \dots y_{m-1})^{n_1} 
(y_2 \dots y_{m-1} )^{n_2} 
\dots
y_{m-1}^{n_m-1} 
\end{array}
\end{equation}
Using Eq. \ref{app2} to replace $1/\Gamma\left( \sum_{i=1}^m (n_i + 1) \right)$ in 
Eq. \ref{app1}, and grouping terms,
\begin{equation}\label{app3}
\begin{array}{ll}
\mathcal{L}^{-1} 
\left\{ 
\Pi_{j=1}^{m} 
\mathcal{L} \left[ f_{j}(t_{j})  \right] 
\right\} 
&= t^{-1+m}
\int_0^1 dy_1 \dots \int_0^1 dy_{m-1} y_1^0 y_2^1 y_3^2 \dots y_{m-1}^{m-2} \times 
\\
& \left( \sum_{n_1=0}^{\infty} f_{1,n_1}  
\frac{t^{n_1}(y_1 \dots y_{m-1})^{n_1}}{\Gamma(n_1+1)} \right)
\\
& \left( \sum_{n_1=0}^{\infty} f_{2,n_2}  
\frac{t^{n_2}(1-y_1)^{n_2} (y_2 \dots y_{m-1})^{n_2}}{\Gamma(n_2+1)} \right)
\\
& \dots 
\\
& \left( \sum_{n_m=0}^{\infty} f_{m,n_m}  
\frac{t^{n_m}(1-y_{m-1})^{n_m}}{\Gamma(n_m+1)} \right)
\end{array}
\end{equation}
The $m$ Taylor series can now be re-summed to give, 
\begin{equation}\label{genSoln}
\begin{array}{c}
\mathcal{L}^{-1} 
\left\{ 
\Pi_{j=1}^{m} 
\mathcal{L} \left[ f_{j}(t_{j})  \right] 
\right\} 
= t^{-1+m}
\int_0^1 dy_1 \dots \int_0^1 dy_{m-1} y_1^0 y_2^1 y_3^2 \dots y_{m-1}^{m-2} \times 
\\
f_1 \left( t  y_1 \dots y_{m-1} \right) 
f_2 \left( t (1-y_1)(y_2 \dots y_{m-1}) \right)
f_3 \left( t (1-y_2)(y_3 \dots y_{m-1}) \right)
\dots 
f_m \left( t(1-y_{m-1}) \right) 
\end{array}
\end{equation}
For example, taking $m=2$ gives, 
\begin{equation}
f(t) = t \int_0^1 dy_1 f_1(ty_1) f_2(t(1-y_1))
\end{equation}
as we could have got from the convolution formula after a simple change of variables. 
Eq. \ref{genSoln} might equivalently be regarded as a generalisation of a 
Schwinger/Feynman parameterisation, with,
\begin{equation}
\begin{array}{ll}
\Pi_{j=1}^m g_j(s) &= 
\int_0^1 dy_1 \dots \int_0^1 dy_{m-1} y_1^0 y_2^1 y_3^2 \dots y_{m-1}^{m-2} \times 
\\
&\mathcal{L} 
\left[
t^{m-1} 
\mathcal{L}^{-1} \left\{ g_1(s) \right\}(t y_1 \dots y_{m-1} )
\right.
\\
&\mathcal{L}^{-1} \left\{ g_2(s) \right\}(t (1-y_1)y_2 \dots y_{m-1} )
\\
&\dots 
\\
&\left. 
\mathcal{L}^{-1} \left\{ g_m(s) \right\}(t (1-y_{m-1}))
\right]
\end{array}
\end{equation}
For example, taking $g_j(s) = 1/(s+a_j)^{p_j}$ and noting that 
$\mathcal{L}^{-1} \left\{ 1/(s+a_j)^{p_j} \right\} = t^{p_j -1} e^{-a_j t}/\Gamma(p_j)$, 
then we get, 
\begin{equation}\label{ident1} 
\begin{array}{ll}
\Pi_{j=1}^m \frac{1}{ (s+a_j)^{p_j} } 
&= 
\frac{\Gamma\left( \sum_{i=1}^m p_i \right)}{\Pi_{i=1}^m \Gamma(p_i) }
\int_0^1dy_1 \dots \int_0^1 dy_{m-1} 
y_1^0 y_2^1 \dots y_{m-1}^{n_m-1} 
\\
&(y_1 \dots y_{m-1})^{p_1 -1} 
\left( (1-y_1) y_2 \dots y_{m-1} \right)^{p_2 -1} 
\dots
\\
&(y_{m-1}(1-y_{m-2}))^{p_{m-1}-1} 
(1-y_{m-1})^{p_m -1} 
\\
&\frac{1}{ \left[ s + (a_1y_1 \dots y_{m-1} 
	               +  a_2(1-y_1)y_2 \dots y_{m-1} 
	               +  \dots 
	               +  a_m(1-y_{m-1}) \right]^{\sum_{i=1}^m p_i } }
\end{array}
\end{equation}
Taking $s=0$, $m=2$, and $p_j=0$ for all $j$, gives the most well-known form, with [27], 
\begin{equation}
\frac{1}{a_1a_2} = \int_0^1 \frac{dy_1}{(a_2y_1+(1-y_1)a_1)^2}  
\end{equation}
The identity Eq. \ref{ident1} can be confirmed by writing the denominator as $(A_m)^m$, with, 
\begin{equation}
A_m = 
\left( a_m(1-y_{m-1}) + y_{m-1} A_{m-1}(y_1, \dots, y_{m-2}) \right) 
\mbox{  and  } A_1=a_1
\end{equation}
and integrating with respect to each of $y_{m-1}$ to $y_1$ in turn. 
For example, using the substitution $u=y_{m-1}/(1+\alpha_{m-1} y_{m-1})$ with 
$\alpha_{m-1}=(A_{m-1}-a_{m})/a_{m}$ and integrating between $u=0$ and 
$u=1/(1+\alpha_{m-1})$, the 
integrand becomes $(1/a_{m})(1/{A_{m-1}}^m)$. 
Repeating this for $y_{m-1}$ to $y_1$ confirms the identity. 

\end{document}